\documentclass[twocolumn,prb]{revtex4}
\usepackage{amsmath}
\usepackage{graphicx}
\usepackage{times}
\usepackage{multirow}
\usepackage{booktabs}

\begin{document}

\title{Molecular Dynamics simulations of concentrated aqueous electrolyte solutions}
\author{Carles Calero}\email{ccalero@icmab.es} 
\author{Jordi Faraudo} 
\affiliation{Institut de Ci\`encia de Materials de Barcelona (ICMAB-CSIC), Campus de la UAB, E-08193 Bellaterra, Spain}
\author{Marcel Aguilella-Arzo}
\affiliation{Biophysics Group, Department of Physics, Universitat Jaume I, 12080 Castell\'o, Spain.}
\begin{abstract}
Transport properties of concentrated electrolytes have been analyzed using classical molecular dynamics simulations with the algorithms and parameters typical of simulations describing complex electrokinetic phenomena. The electrical conductivity and transport numbers of electrolytes containing monovalent (KCl), divalent (MgCl$_2$), a mixture of both (KCl + MgCl$_2$), and trivalent (LaCl$_3$) cations have been obtained from simulations of the electrolytes in electric fields of different magnitude. The results obtained for different simulation parameters have been discussed and compared with experimental measurements of our own and from the literature. The electroosmotic flow of water molecules induced by the ionic current in the different cases has been calculated and interpreted with the help of the hydration properties extracted from the simulations. 
\end{abstract}

\date{\today }
\maketitle

%\begin{figure}[ht!]
%   \begin{center}
%      \includegraphics[angle=-90,width=13.5cm]{Iratio.eps}
%      \caption{Current ratio Icat$/$Ian}
%      \label{fig2}
%   \end{center}
% \end{figure}
\section{Introduction}

Understanding the transport properties of electrolytes in aqueous solution is important in a wide range of electrokinetic phenomena such as streaming current experiments \cite{Heyden2006}, ionic transport in biological and syntetic nanochannels \cite{Alcaraz2009,TravessetSilica} or colloidal electrophoresis \cite{Jonsson2005,MartinMolina2008}. 

Traditionally, electrokinetic and ionic transport phenomena have been described using primitive models in which the solvent is approximated as a continuum of dielectric constant $\epsilon$. Although such models provide an accurate description of a wide variety of phenomena, they fail for cases in which the discrete nature of the solvent plays a fundamental role, or to describe phenomena related to specific hydration of ions. 

An alternative theoretical approach to understand electrokinetic phenomena which circumvents the limitations of primitive models is the use of molecular dynamics simulations of ions in explicit solvent. Due to recent improvements in algorithms and computer power, all-atom molecular dynamics simulatios studies of electrokinetic phenomena in some realistic systems are now possible \cite{Aksimentiev, Lorenz}. From such atomistic descriptions, one can elucidate the microscopic mechanisms responsible for macroscopic measurable phenomena. For example, Aksimentiev and Schulten studied at atomic detail the permeation of individual ions through the transmembrane channel $\alpha$-Hemolysin with the help of molecular dynamics simulations \cite{Aksimentiev}. This approach has also been employed to understand from a microscopic perspective the electrophoresis of DNA immersed in multivalent ionic solutions when an external electric field is applied \cite{AksimentievADN}, the transport of monovalent and divalent ions through polymeric and silica nanopores  \cite{Cruz, CruzSilica}, and the ionic selectivity of the OmpF porin biological nanochannel \cite{MarcelOmpF}. 

The realistic atomic description of such electrokinetic phenomena involves complex systems containing large number of particles. To be able to cope with such systems using molecular dynamics simulations, the use of algorithms which enhance the computer performance are required. A molecular dynamics simulation package which has been proven to be very successful in the description of biological molecules and elecrokinetic phenomena is NAMD2 \cite{NAMD}. To accelerate the calculations of such large systems, in NAMD2 the temperature T of the system is controlled by using the Langevin thermostat instead of the more rigorous but much more computationally demanding Nos\'e-Hoover thermostat. Also to speed up the simulations, NAMD2 is usually employed with a multiple time step: a basic timestep for short range interactions and a longer one for the evaluation of k-space contribution to the long range electrostatic forces in the PME technique. 

In this paper we study, under different conditions, the transport properties obtained from Molecular Dynamics simulations of electrolyte solutions using the same algorithms and conditions employed in simulations of complex systems. In particular, we focus our analysis on the transport properties of a monovalent electrolyte (KCl), a divalent electrolyte (MgCl$_2$),  the mixture of both (KCl $+$ MgCl$_2$) and a trivalent electrolyte (LaCl$_3$)  when an external electric field is applied. While the use of the KCl electrolyte is standard in molecular dynamics studies and its transport properties have been briefly analyzed from this perspective before \cite{Aksimentiev}, the reliability of such simulations methods in dealing with transport properties of multivalent electrolytes have not been tested in detail in spite of the rich phenomenology that they originate \cite{Heyden2006, Alcaraz2009,Gelbart2000,Besteman2007} and their use in the simulation of electrokinetic phenomena involving divalent and trivalent ions \cite{Jin, Luan, Craig, AksimentievADN}.

\section{Methods}
\subsection{Simulation Methods}
We have studied the transport properties of different electrolytes by performing molecular dynamics simulations under external electric fields of different magnitudes. The systems considered were ionic solutions of KCl, MgCl$_2$, a mixture of KCl and MgCl$_2$, and LaCl$_3$, see Table I. To study the effect of the system size on the ionic transport properties of bulk electrolytes extracted from molecular dynamics simulations, two cubic simulation boxes of different size were used, $L \simeq 4$nm and $L \simeq 8$nm.    

All simulations have been carried out with the molecular dynamics simulation package NAMD2\cite{NAMD}, since it is widely employed in the simulation of biological macromolecules. Water was described using the TIP3P water model as implemented by the CHARMM force field. The ions were modeled as charged Lennard-Jones particles with parameters given by the CHARMM force field for K$^+$, Mg$^{2+}$ and for Cl$^-$, while the Lennard-Jones parameters for La$^{3+}$ were taken from Ref. \cite{LJ_La}. 

\begin{table*}
\caption{Simulation parameters for the simulations performed of the different electrolytes: cubic box edge length (L), damping constant of the Langevin thermostat ($\tau_{\text{Lan}}$), number of K$^+$ ions, number of Mg$^{2+}$ ions, number of La$^{3+}$ ions, number of Cl$^-$ ions, number of water molecules, simulation time, external electric field applied, and potential difference between the edges of the cubic box along the direction of the electric field. }
\label{Table:Table1}
\centering
    \begin{tabular}{ |l||l|l|l|l|l|l|l|l|l|l|}
    \hline
    Electrolyte & L (nm) & $\tau_{\text{Lan}}$ (ps$^{-1}$) &Num.  & Num.  & Num.  & Num.  & Num.  & Simulation  & Electric  &  $\Delta$V (mV)\\
		&	&  	 &K$^+$	& Mg$^{2+}$ &	La$^{3+}$ & Cl$^-$	&H$_2$O		& time (ns)& field (mV/nm)&	\\ \hline
 
    1M KCl & 3.88 & 1.0 & 37 & 0 & 0 & 37 & 1907 & 21.0 & 14.2 & 55 \\
	   & 3.88 & 1.0 &37 & 0 & 0 & 37 & 1907 & 9.0 & 26.0 & 109\\
	   & 3.88 & 1.0 &37 & 0 & 0 &  37 & 1907 & 9.0 & 43.3 & 168\\
	   & 3.88 & 1.0 &37 & 0 & 0 &  37 & 1907 & 9.0 & 86.6 & 336\\ \cline{2-11}

	   & 7.82 & 1.0 & 306 & 0 & 0 &  306 &  15774 & 9.0 & 14.2 & 111\\
	   & 7.82 &1.0 & 306 & 0 & 0 &  306 &  15774 & 9.0 & 26.0 & 203\\
	   & 7.82 & 1.0 &306 & 0 & 0 &  306 &  15774 & 9.0 & 43.3 & 339\\
	   & 7.82 & 1.0 &306 & 0 & 0 &  306 &  15774 & 9.0 & 86.6 & 677\\ \hline

    1M MgCl$_2$ & 3.82 & 1.0 &0 & 37 & 0 &  74 & 1870 & 9.0 & 14.2 & 54\\
		& 3.82 & 1.0 &0 & 37 & 0 &  74 & 1870 & 9.0 & 26.0 & 99\\
		& 3.82 & 1.0 &0 & 37 & 0 &  74 & 1870 & 9.0 & 43.3 & 165\\
		& 3.82 & 1.0 &0 & 37 & 0 &  74 & 1870 & 9.0 & 86.6 & 331\\ \cline{2-11}

	        & 7.73 & 1.0 &0 & 306 & 0 &  612 & 15468 & 9.0 & 14.2 & 110\\
	        & 7.73 & 1.0 &0 & 306 & 0 &  612 & 15468 & 9.0 & 26.0 & 201\\
	        & 7.73 & 1.0 &0 & 306 & 0 &  612 & 15468 & 9.0 & 43.3 & 335\\
	        & 7.73 & 1.0 &0 & 306 & 0 &  612 & 15468 & 9.0 & 86.6 & 669\\ \cline{2-11}

	        & 7.73 & 0.2 &0 & 306 & 0 &  612 & 15468 & 9.0 & 14.2 & 110\\
	        & 7.73 & 0.2 &0 & 306 & 0 &  612 & 15468 & 9.0 & 26.0 & 201\\
	        & 7.73 & 0.2 &0 & 306 & 0 &  612 & 15468 & 9.0 & 43.3 & 335\\
	        & 7.73 & 0.2 &0 & 306 & 0 &  612 & 15468 & 9.0 & 86.6 & 669\\ \hline

    0.11M MgCl$_2$ & 7.83 &  1.0 & 0 & 31 & 0 &  62 & 16293 & 9.0 & 43.3 & 339\\
    0.33M MgCl$_2$ & 7.80 & 1.0 & 0 & 93 & 0 &  186 & 16107 & 9.0 & 43.3 & 338\\
    0.54M MgCl$_2$ & 7.79 & 1.0 & 0 & 155 & 0 &  310 & 15921 & 9.0 & 43.3 & 337\\ \hline
    1M MgCl$_2$ & 7.72 & 1.0 & 306 & 306 & 0 &  918 & 14856 & 9.0 & 14.2 & 110\\
		 + 1MKCl & 7.72 & 1.0 & 306 & 306 & 0 &  918 & 14856 & 9.0 & 26.0 & 201\\
			& 7.72 & 1.0 & 306 & 306 & 0 &  918 & 14856 & 9.0 & 43.3 & 334\\
			& 7.72 & 1.0 & 306 & 306 & 0 &  918 & 14856 & 9.0 & 86.6 & 668\\ \hline
    1M LaCl$_3$ & 7.71 & 1.0 & 0 & 0 & 308 & 924 & 15154 & 13.19 & 14.2 & 110\\
		& 7.71 & 1.0 & 0 & 0 & 308 & 924 & 15154 & 10.87 & 26.0 & 201\\
		& 7.71 & 1.0 & 0 & 0 & 308 & 924 & 15154 & 10.12 & 43.3 & 334\\
		& 7.71 & 1.0 & 0 & 0 & 308 & 924 & 15154 & 10.52 & 86.6 & 668\\ \hline

    \end{tabular}
\end{table*}

\begin{figure}[ht!]
   \begin{center}
      \includegraphics[angle=0,width=7cm]{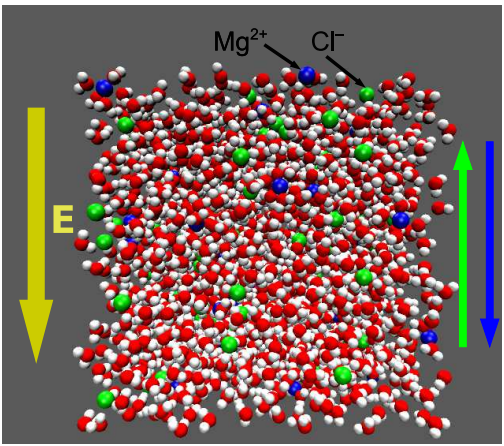}
      \caption{Snapshot of a simulation of the 1M MgCl$_2$ electrolyte with simulation box size of $L = 3.82$ nm. Chloride ions are represented in green, magnesium ions in blue, and water molecules in red (oxygen) and white (hydrogen). The yellow arrow points in the direction of the applied electric field, the green arrow in the direction of the flow of chloride ions (anions), and the blue arrow in the direction of the flow of magnesium ions (cations).}
      \label{Fig:diagram}
   \end{center}
 \end{figure}

The initial configuration of the simulation was constructed as follows. The ions (K$^+$, Mg$^{2+}$ and Cl$^-$) were inserted at random positions by employing the {\it AutoIonize} plugin of the Visual Molecular Dynamics (VMD) software package \cite{VMD} inside a cube of preequilibrated TIP3P water molecules, obtained with the help of the {\it Solvate} plugin of VMD (see Fig. \ref{Fig:diagram} for a diagram of the system). The faces of the cube are parallel to the XY, XZ, and YZ planes. 
In all simulations we employed periodic boundary conditions in all directions. Lennard-Jones interactions were computed using a smooth (10-12 \AA) cutoff, as it is customary done in NAMD2 simulations \cite{Aksimentiev, AksimentievADN}. The electrostatic interactions were calculated using the particle-mesh Ewald (PME) method with a precision of $10^{-6}$ using a $128\times128\times128$ grid and a $12$ \AA cutoff for the real space calculation. These are common parameters used to simulate complex and large biological macromolecules \cite{Aksimentiev}.

For each case, the equilibration procedure consisted of 50000 steps of energy minimization, a 1ns run in the NpT ensemble (with $p= 1$atm and $T = 296$K) followed by a 1ns run in the NVT ensemble ($T = 296$K). Production runs in the NVT ensemble were performed in the presence of different electric fields to induce electromigration of the ions in solution. The NpT simulation runs were performed employing a combination of the Nos\'e-Hoover constant pressure method with the piston fluctuation control implemented using Langevin dynamics as implemented in NAMD2 ($p = 1$atm, period of the oscillations of $0.1$ps and relaxation constant of $0.05$ ps). As mentioned above, to speed up calculations the NVT runs were carried out by applying Langevin dynamics, with parameters (also in the NpT run) $T=296$K and a damping coefficient of $1$ps$^{-1}$ (using $0.2$ps$^{-1}$ to test its effect in the dynamics for some cases specified later). Langevin forces were applied to all atoms except for hydrogens, which thermalize through interactions with the rest of the system. 

In all cases, the equations of motion were solved using a multiple time step in order to speed up the simulations. A basic time step of 2fs was used for the evaluation of short range interactions and a longer time step of 4fs for the calculations of the k-space contribution to the long range electrostatic forces in the PME technique.

In the production runs, the instantaneous current induced by the external electric field applied along the Z-direction is calculated with the help of  \cite{Aksimentiev}
\begin{equation}
\label{Ieq}
 I(t) = \frac{1}{\Delta t L}\sum_{i =1}^{N}q_i\left[ z_i(t+\Delta t) - z_i(t)\right],
\end{equation}
where $z_i$ and $q_i$ are the Z-coordinate and the charge of atom $i$, respectively. $L$ is the size of the simulation box and $\Delta t$ is the time interval employed to record data, which was chosen to be $\Delta t = 10$ps. The average current is computed by linearly fitting the cumulative current that is obtained by integration of the instantaneous current. To ensure constistency of the results, the current was also computed by counting the flux of ions crossing a plane perpendicular to the direction of the electric field and located in the center of the simulation box.  The conductivity $\kappa$ of the solution is defined by:
\begin{equation}
\label{conductivity}
 \frac{I}{A}=\kappa E,
\end{equation}
where $A= L^2$ is the cross section area perpendicular to the electric field $E$. In order to obtain the conductivity of the electrolyte, we have performed simulations at different electric fields $E$ and calculated the current $I$ induced by them. In all cases, an ohmic behaviour has been observed (i.e. consistent with a linear relation between $I$ and $E$), and a linear fit of the data to Eq.\,(\ref{conductivity}) gives the value of the conductivity $\kappa$. 

The electromigration of ions in aqueous solutions as a result of the application of an external electric field is often accompanied by a net flow of water, the so-called electroosmotic flow. The electroosmotic flow has been evaluated by keeping track, every $\Delta t = 10$ps, of the accumulated number of water molecules crossing a plane perpendicular to the electric field. Crossings of such plane in the direction of the electric field are counted as positive, whereas crossings in the opposite direction are counted as negative. Hence, the sign of the overall flow gives the direction of the electroosmotic flow with respect to the elecric field direction. As reported before \cite{Aksimentiev}, the simulations might give a drift of the whole system which is unphysical since no net force is applied to the system (the electrolyte is globally electrically neutral). To avoid such spurious effects, the computation of the electroosmotic flow has been done in the frame of reference of the center of mass of the whole system. 

%The electric current flowing in an electrolyte solution is caused by the motion of ions of opposite charge moving in opposite directions under the applied field. The fraction of the total current induced by each ion type defines its transport number, which is in general a function of the electrolyte concentration. Due to the global electroneutrality of the system, the total electric current flowing through an electrolyte as a result of the application of an external electric field is independent of the frame of reference in which it is measured. Transport numbers, however, are frame dependent, and their computation must be done carefully.  A proper account of transport numbers in electrolyte solutions is important to describe phenomena in other more complex systems in which not only the total elctric current but also the flow of each type of ion is relevant (e.g. in the study of the selectivity of ionic channels). Experimentally, the relevant frame of reference in which data is provided is the frame of reference of the moving fluid. Thus, transport numbers will be computed from the simulations results in such frame of reference to facilitate the comparison  with experimental values.

The electric current flowing in an electrolyte solution is caused by the motion of anions and cations moving in opposite directions under the applied field. The fraction of the total current induced by each ion type defines its transport number, which is in general a function of the electrolyte concentration. The fraction of electrical current carried by cations defines the cationic transport number, t$_+$, and the fraction of the electrical current carried by anions the anionic transport number, t$_-$. A completely equivalent quantity to transport numbers is the ratio of the current induced by anions (anionic current) over the current induced by cations (cationic current), which will be be refered as the ratio of currents throughout the paper. Due to the global electroneutrality of the system, the total electric current flowing through an electrolyte as a result of the application of an external electric field is independent of the frame of reference in which it is measured. The ratio of currents and transport numbers, however, are frame dependent, and their computation must be done carefully. A proper account of the contribution to the total current from every ion in electrolyte solutions is important to describe phenomena in other more complex systems in which not only the total electric current but also the flow of each type of ion is relevant (e.g. in the study of the selectivity of ionic channels).  Experimentally, the relevant frame of reference in which data of transport numbers and ratio of currents is provided is the frame of reference of the moving fluid. Therefore, to facilitate the comparison  with experimental values, transport numbers and ratio of currents will be given in the frame of reference comoving with the fluid.

\subsection{Conductivity measurements}

The electrolyte conductivity measurements were performed using a MeterLab CDM210 (R) conductivity meter, Radiometer Analytical SAS (France). The solutions were prepared using water from a Water Purification System Millipore Simplicity 185. Magnesium Chloride 6-hydrated (MgCl2) and Potassium Chloride (KCl) from Panreac  in all cases following ACS specifications. Weighting of the compounds were done with a Mettler Toledo AB104-S, in the quantities necessary to achieve a 1M concentration. The conductivity measurement included a stirring of the solution with a magnetic stirrer and heater JPSelecta Agimatic-E, with temperature monitoring using a laboratory thermometer at 296.0 K.

%and Lanthanum Chloride (LaCl3) from JT Baker were used,

\section{Results}
\subsection{A test case: Transport properties of 1M KCl}

\begin{figure}[ht!]
   \begin{center}
      \includegraphics[angle=-90,width=8.5cm]{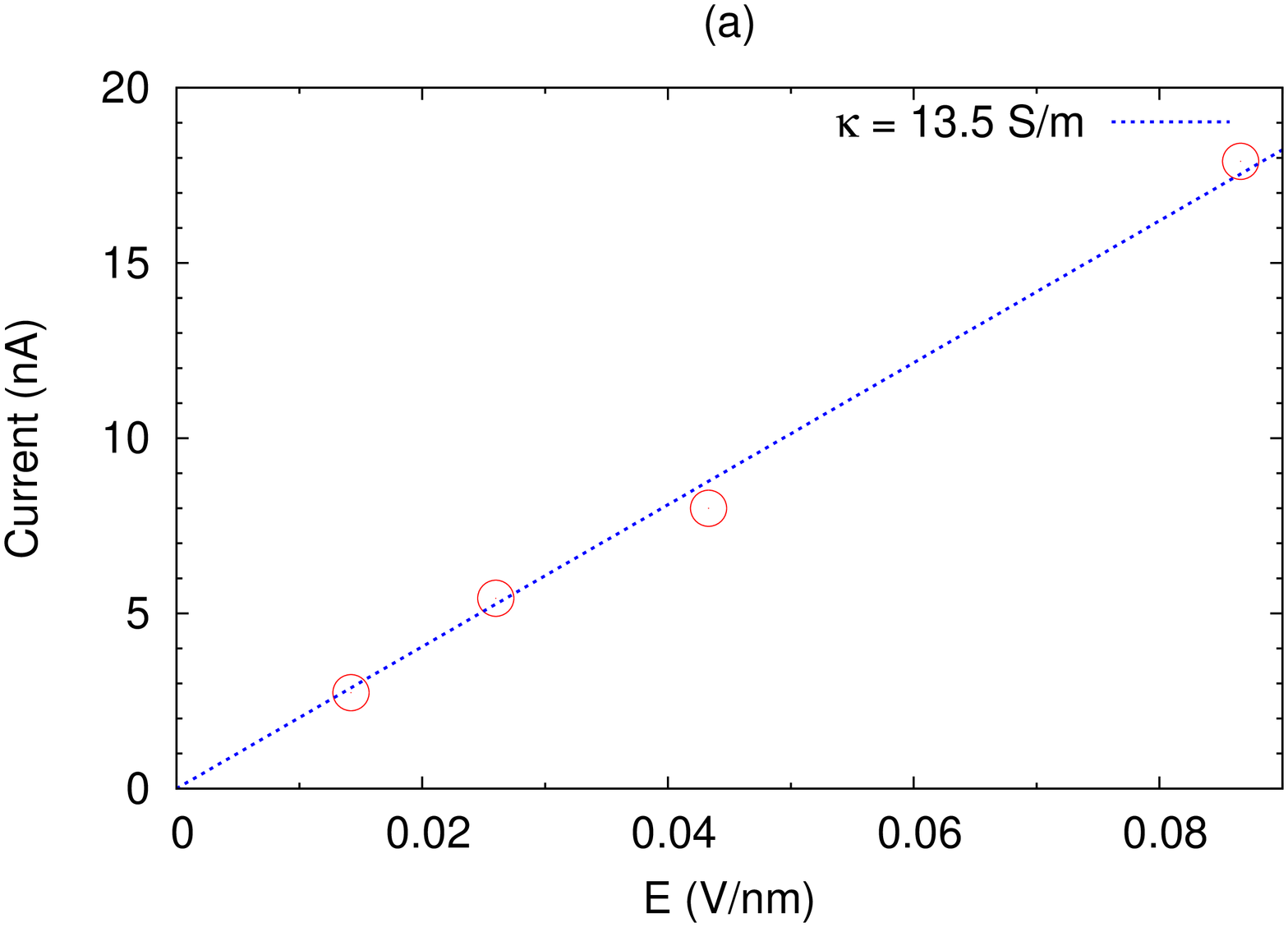}
      \includegraphics[angle=-90,width=8.5cm]{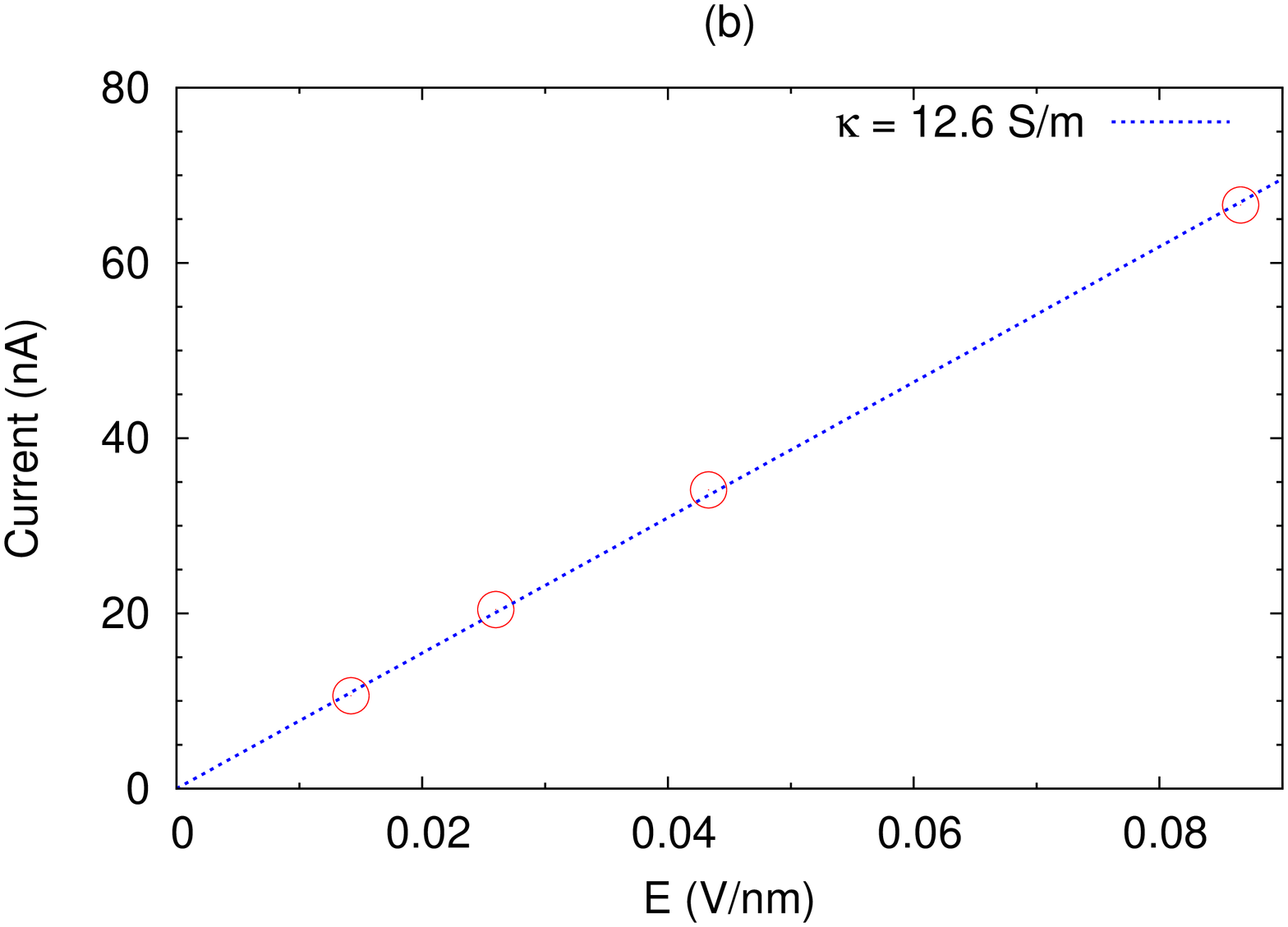}
      \caption{Electric current versus the applied electric field for a 1M KCl electrolyte in simulations using two different simulation box sizes $L$. The points correspond to simulation data computed from Eq.(\ref{Ieq}) and lines are fits of the data to Eq.(\ref{conductivity}) (a) $L = 3.88$ nm (fit gives $\kappa = 13.4$S/m).(b) $L = 7.82$ nm (fit gives $\kappa = 12.6$S/m).}
	\label{Fig:KCl}
   \end{center}
 \end{figure}

The results obtained from molecular dynamics simulations for the electrical properties of the 1M KCl electrolyte  are given in Fig. \ref{Fig:KCl} and Table.\ref{Table:IRatio_KCl}. We have used two different sizes for the simulation box, $L = 3.88$nm and $L = 7.82$nm, to test the dependence of the results on the simulation's box size. As can be seen, the values for the ionic conductivity for the two different box sizes do not differ significantly, $\kappa = 13.4$ S/m for the small box and $\kappa = 12.6$ S/m for the big one. Both results are in good agreement with our measured experimental value $\kappa_{\text{exp}} = 11.24\pm 0.01$ S/m, being the value of the larger system closer to the experimental result as it should be expected. Note that in previous studies of 1M KCl bulk electrolyte (see Ref.\cite{Aksimentiev}), it was argued that a $0.2$ ps$^{-1}$ damping constant is necessary in order to reproduce correctly transport properties. However, our present results were obtained using a Langevin thermostat with a damping constant of $1$ ps$^{-1}$, which is the typical choice for simulations of complex systems in contact with electrolyte (such as protein channels or silica nanochannels \cite{Aksimentiev, MarcelOmpF,CruzSilica}). Our results show that with the standard simulation parameters employed in complex systems the conductivity of the electrolyte KCl is correctly reproduced. 

%We have also calculated the cationic transport numbers of the electrolyte at different values of the external electric field. As explained in the previous section, it is defined as the fraction of the total electrical current caused by the flow of cations in the frame of reference of the moving fluid. The results for the transport numbers are given in Fig. \ref{Fig:TransportNumbersKCl}. As it is shown in the figure, the values calculated from the molecular dynamics simulations are in good agreement with the experimental value of $0.4882$\cite{Stokes}, especially at low electric fields. 

We have also calculated the ratio of the different contributions to the total current from anions and cations (equivalent to the ratio of transport numbers). As explained in the previous section, it is defined as the ratio of the anionic and cationic currents in the frame of reference of the moving fluid. The results for such ratio of currents are given in Table \ref{Table:IRatio_KCl}. As shown in the table, the values calculated from molecular dynamics simulations are in good agreement with the experimental value of $I_{\text{Cl}}/I_\text{{K}} = 1.048$ (equivalent to a value of the cation transport number t$_+ = 0.488$ \cite{Stokes}) for both sizes of the cubic simulation box. 

%The results for such ratio of currents are given in Fig. \ref{Fig:ratioKCl8}. As shown in the figure, the values calculated from the molecular dynamics simulations are in good agreement with the experimental value of $1.048$\cite{Stokes}, especially at low electric fields. 

%\begin{figure}[ht!]
%   \begin{center}
%      \includegraphics[angle=-90,width=8cm]{TransportNumbers_KCl4.eps}
%      \includegraphics[angle=-90,width=8cm]{TransportNumbers_KCl8.eps}
%      \caption{Cationic transport numbers of 1M KCl for different electric fields using two different simulation box sizes $L$. The dots are the results of the calculation from the Molecular Dynamics simulations results for (a)$L = 3.88$, and (b) $L = 7.82$nm. The dashed line gives the experimental value obtained for 1M KCl \cite{Stokes} }
%      \label{Fig:TransportNumbersKCl}
%   \end{center}
% \end{figure}

%\begin{figure}[ht!]
%   \begin{center}
%      \includegraphics[angle=-90,width=8cm]{Iratiow_KCl4.eps}
%      \includegraphics[angle=-90,width=8cm]{Iratiow_KCl8.eps}
%      \caption{ Ratio of the anionic current over the cationic current for different electric fields, as calculated in the frame of reference of the moving fluid. The dots are the results of the calculation from the Molecular Dynamics simulations results for two different sizes of the cubic simulation box: (a)$L = 3.88$, and (b) $L = 7.82$nm. The dashed line gives the experimental value obtained for 1M KCl \cite{Stokes}.}
%      \label{Fig:Iratio_KCl}
%   \end{center}
% \end{figure}

\begin{table}
\caption{Ionic contributions to the total current, ratio of the anionic current over the cationic current and cationic transport numbers for different values of the applied electric field in the case of a 1M KCl electrolyte. All these quantities are calculated in the frame of reference of the moving fluid.}
\label{Table:IRatio_KCl}
\centering
%\left
%\begin{tabular}{l|c c c | c c}
\begin{tabular}{|c|c|c|c|c|c|}
\hline \hline
 System Size & Electric field (mV/nm) & I$_{\text{Cl}}$ (nA) & I$_{\text{K}}$ (nA) & I$_{\text{Cl}}$/I$_{\text{K}}$ & t$_+$ \\
\hline
\multicolumn{1}{|c|}{\multirow{4}{*}{L = 3.88 nm}} &
 14.2 & 1.43 & 1.39 & 1.03 &  0.49 \\
& 25.98 & 2.80 & 2.59 & 1.08 & 0.48 \\
& 43.3 & 4.28 & 3.83 & 1.12 & 0.47\\
& 86.6 & 8.42 & 8.22 & 1.02 & 0.49 \\
\hline
\multicolumn{1}{|c|}{\multirow{4}{*}{L = 7.82 nm}} &
 14.2 & 5.49 & 5.14 & 1.07 &  0.48 \\
& 25.98 & 10.35 & 10.11 & 1.02 & 0.49 \\
& 43.3 & 17.02 & 17.16 & 1.00 & 0.50\\
& 86.6 & 33.57 & 33.17 & 1.01 & 0.49 \\
\hline

\end{tabular} 
\end{table}

\subsubsection{Electoosmotic flow}
\begin{figure}[ht!]
   \begin{center}
      \includegraphics[angle=-90,width=8cm]{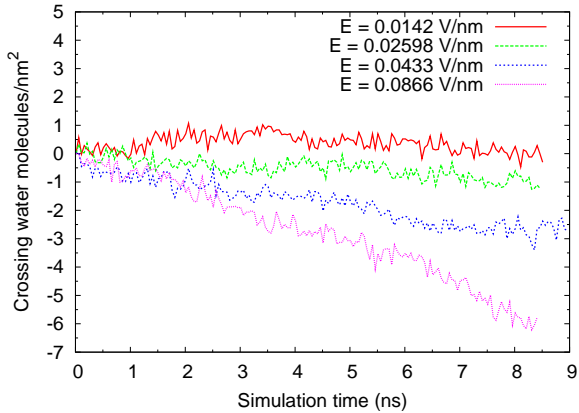}
      \caption{Accumulated number of water molecules per unit area crossing a plane perpendicular to the applied electric field versus simulation time for the 1M KCl electrolyte. Different lines represent the flow of water for different values of the electric field.}
      \label{Fig:water_flowKCl}
   \end{center}
 \end{figure}

In addition to ionic currents, we also observe a net flow of water molecules. In Fig. \ref{Fig:water_flowKCl}, the accumulated number of water molecules per unit area crossing a plane perpendicular to the electric field is represented as a function of simulation time. A linear fit of these magnitudes provides the flux of water molecules for every value of the electric field, as shown in Table \ref{Table:Water_Flow}. As shown in Fig. \ref{Fig:water_flowKCl} and Table \ref{Table:Water_Flow}, the electroosmotic flow for the 1M KCl electrolyte is in the direction opposite to the external field, that is, in the direction of the flow of chloride ions.

\begin{table}
\caption{Flux of water molecules for the different electrolytes and different values of the external electric field.}
\label{Table:Water_Flow}
\centering
%\left
%\begin{tabular}{l|c c c | c c}
    \begin{tabular}{|c|c|c|}
\hline
 Electrolyte & Electric field (mV/nm) & Water flux (nm$^{-2}$ns$^{-1}$) \\
\hline
 1M KCl & 14.2 & 0.12 $\pm$ 0.13\\
 	& 26.0 &  -0.13 $\pm$ 0.04\\
	& 43.3 &  -0.43 $\pm$ 0.14\\
	& 86.6 &  -0.64 $\pm$ 0.07\\
\hline
 1M MgCl$_2$ & 14.2 & 0.76 $\pm$ 0.04\\
 	& 26.0 &  1.47 $\pm$ 0.05\\
	& 43.3 &  2.30 $\pm$ 0.14\\
	& 86.6 &  5.11 $\pm$ 0.09\\
\hline
 1M KCl $+$ 1M MgCl$_2$ & 14.2 & 0.43 $\pm$ 0.15\\
 	& 26.0 &  0.88 $\pm$ 0.06\\
	& 43.3 &  1.31 $\pm$ 0.07\\
	& 86.6 &  2.82 $\pm$ 0.23\\
\hline
 1M LaCl$_3$ & 14.2 & -0.25 $\pm$ 0.11\\
 	& 26.0 &  -0.28 $\pm$ 0.25\\
	& 43.3 &  -0.33 $\pm$ 0.08\\
	& 86.6 &  -0.89 $\pm$ 0.24\\

\hline
    
    \end{tabular} 
\end{table}
\subsection{Transport properties of 1M MgCl$_2$}

\begin{figure}[ht!]
   \begin{center}
      \includegraphics[angle=-90,width=8.5cm]{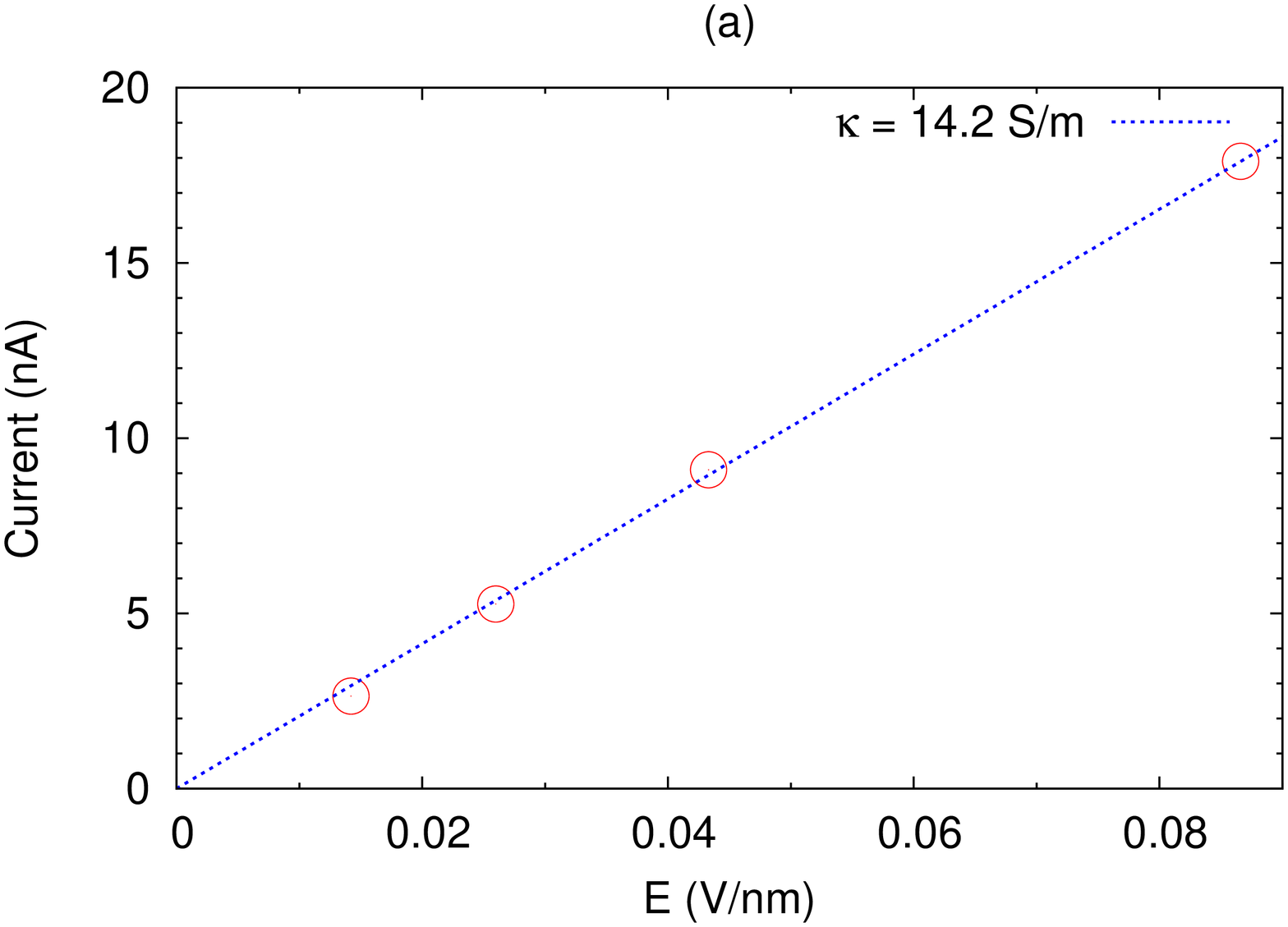}
      \includegraphics[angle=-90,width=8.5cm]{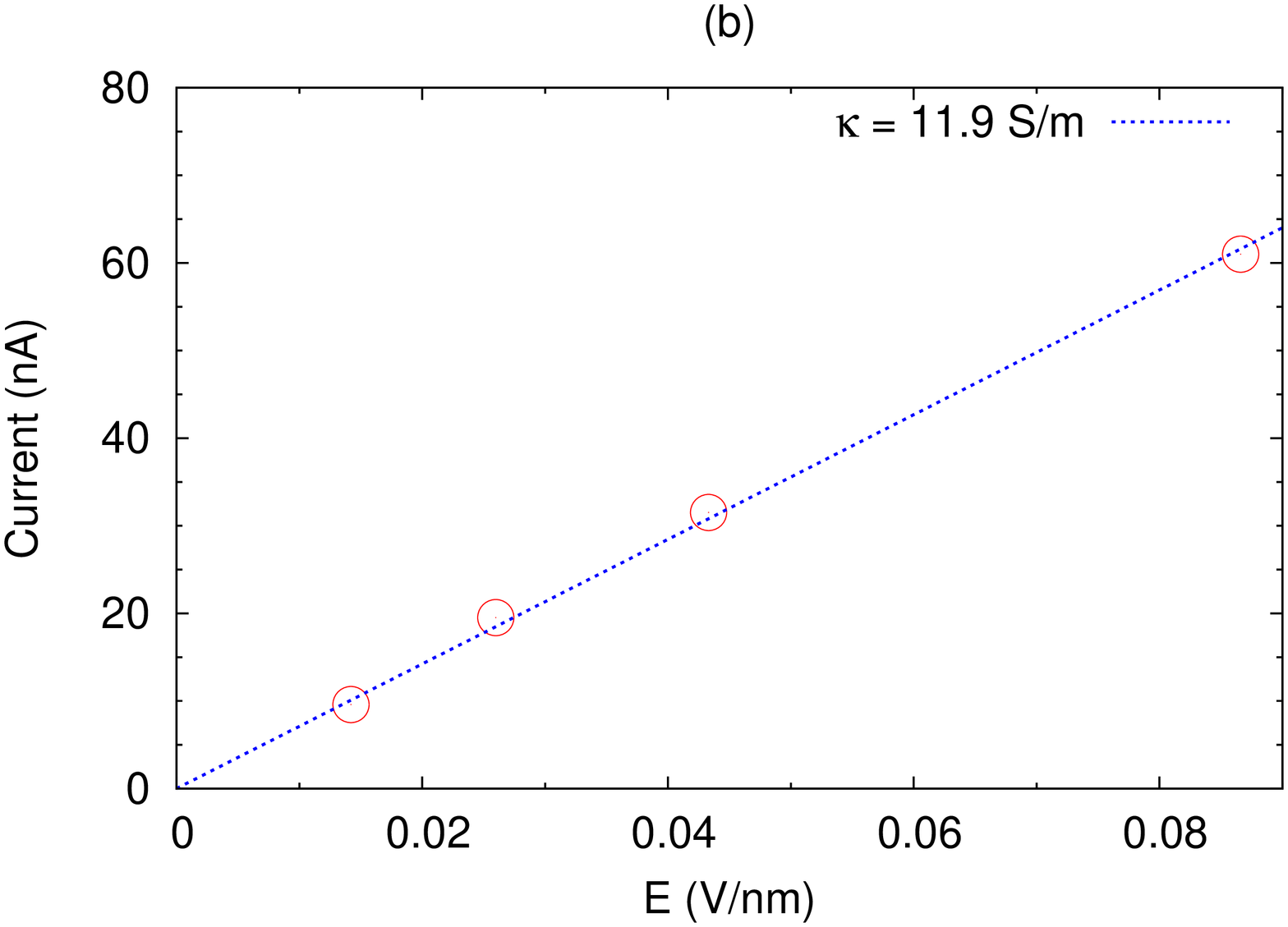}
      \caption{Electric current versus applied electric field for a 1M MgCl$_2$ electrolyte in simulations using two different simulation box sizes $L$. The points correspond to simulation data computed from Eq.(\ref{Ieq}) and lines are fits of the data to Eq.(\ref{conductivity}) (a) $L = 3.82$nm size cubic simulation box (fit gives $\kappa = 14.2$S/m). (b) $L = 7.73$nm (fit gives $\kappa = 11.9$S/m)}
      \label{Fig:MgCl}
   \end{center}
 \end{figure}

A similar procedure has been followed to obtain the transport properties of a 1M MgCl$_2$ electrolyte. The electric current induced by the electromigration of ions was obtained for different values of the electric field using two different sizes of a cubic simulation box, $L = 3.82$ nm and $L = 7.73$ nm. The results of the simulations are given in Figs. \ref{Fig:MgCl} and Table \ref{Table:IRatio_MgCl2}. The values for the ionic conductivity of the two different box sizes do not differ significantly, being $\kappa = 14.2$ S/m for the small box and $\kappa = 11.9$ S/m for the big one. Both results agree well with experimental values, being the result from the larger simulation box much more accurate. According to Phang and Stokes\cite{Phang1980}, $\kappa_{\text{exp}} = 11.4$ S/m at [MgCl$_2$]=0.9674 M and $T=298.15$K. A recent critical review\cite{Miller1984} provides a sligthly larger value of  $\kappa_{\text{exp}} = 11.6$ S/m at [MgCl$_2$]=1 M and $T=25^o$C. Our own measurements give a value of $\kappa_{\text{exp}} = 11.92$ S/m.

\begin{table}
\caption{Ionic contributions to the total current, ratio of the anionic current over the cationic current and cationic transport numbers for different values of the applied electric field in the case of a 1M MgCl$_2$ electrolyte. All these quantities are calculated in the frame of reference of the moving fluid.}
\label{Table:IRatio_MgCl2}
\centering
%\left
%\begin{tabular}{l|c c c | c c}
\begin{tabular}{|c|c|c|c|c|c|}
\hline \hline
 System  & Electric  & I$_{\text{Cl}}$ (nA) & I$_{\text{Mg}}$ (nA) & I$_{\text{Cl}}$/I$_{\text{Mg}}$ & t$_+$ \\
 Size 	& field (mV/nm) &		& 	&		&		\\ 
\hline
\multicolumn{1}{|c|}{\multirow{4}{*}{L = 3.82 nm}} &
 14.2 & 1.60 & 0.72 & 2.23 &  0.31 \\
& 25.98 & 3.09 & 2.22 & 1.39 & 0.42 \\
& 43.3 & 5.63 & 3.56 & 1.58 & 0.39\\
& 86.6 & 10.52 & 7.42 & 1.42 & 0.41 \\
\hline
\multicolumn{1}{|c|}{\multirow{4}{*}{L = 7.73 nm}} &
 14.2 & 5.69 & 4.13 & 1.38 &  0.42 \\
& 25.98 & 11.31 & 8.24 & 1.37 & 0.42 \\
& 43.3 & 18.05 & 13.66 & 1.32 & 0.43\\
& 86.6 & 35.06 & 26.30 & 1.33 & 0.43 \\
\hline

\end{tabular} 
\end{table}

The ratio between the different contributions of anions and cations to the total current has been evaluated. The results for such ratio of currents are given in Table \ref{Table:IRatio_MgCl2}. These results emphasize the need for using a large enough simulation box. The values for the ratio of currents and transport numbers obtained for the cubic simulation box of size $L = 3.82$nm exhibit a spurious dependence on the applied electric field. For the larger simulation box, the results are not field dependent, as should be expected. Such results show that the anionic contribution to the current is significantly larger than the cationic contribution (being a factor of 1.4 between them). In our simulations these differences in the anionic and cationic contributions to the current can be attributed, to a large extent, to differences in diffusion coefficients between both ions. In order to disentangle the diffusional and correlation contributions to the transport number, we have evaluated the translational diffusion coefficient of each ion by computing the mean square displacement of each ion in a 2ns long NVE simulation run with no external field applied. The results of $D_{\text{Mg}}=0.95\times 10^{-5}$ cm$^2$/s and $D_{\text{Cl}}=1.69\times 10^{-5}$ cm$^2$/s for the diffusion coefficients of Mg$^{2+}$ and Cl$^-$, respectively, lead to a ratio between the diffusion coefficients of $D_{\text{Cl}}/D_{\text{Mg}}=1.8$.

\begin{figure}[ht!]
   \begin{center}
      \includegraphics[angle=-90,width=8.5cm]{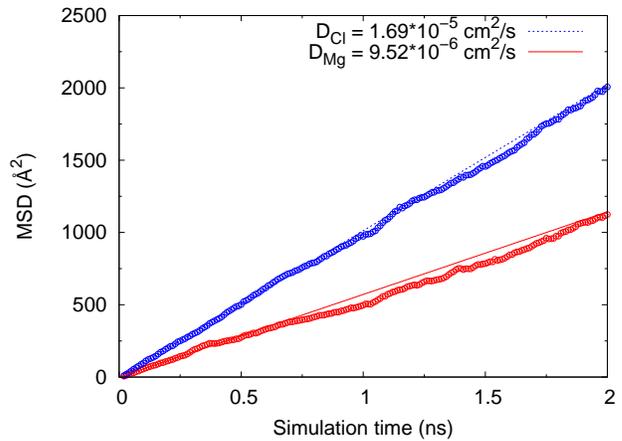}
      \caption{Diffusion of Mg$^{2+}$ and Cl$^-$ ions in 1 M aqueous solution as obtained from 2ns NVE simulations with a cubic simulation box of $ L = 7.73$nm.}
      \label{Fig:Diffusion}
   \end{center}
 \end{figure}

Experimentally, both diffusion coefficients and transport numbers can be obtained, so we can compare both simulation results with experimental data. Using NMR, Struis et al.\cite{Struis1989} obtained $D_{\text{Mg}}=0.93 \times 10^{-5}$ cm$^2$/s and $D_{\text{Cl}}=2.94 \times10^{-5}$ cm$^2$/s (at 25$^o$C and 0.985 mol/Kg concentration), so the experimental ratio is $D_{\text{Cl}}/D_{\text{Mg}}=3.16$. The simulations reported here reproduce with good accuracy the diffusion coefficient for Mg$^{2+}$ but the diffusion coefficient for Cl$^-$ is greatly underestimated. Concerning transport numbers, electrochemical measurements\cite{Phang1980,Miller1984} give a transport number for cations around 0.3, so the experimental ratio between anionic and cationic currents is $\simeq 2.3$. The difference between the ratio of transport numbers obtained by electrochemical methods and the ratio between diffusion coefficients obtained by NMR can be interpreted in several ways. First of all, thermodynamic arguments\cite{Miller1966} show that some experimental procedures mixed up different reference frames, so the electrochemical results have to be interpreted with caution. If the difference between both ratios is indeed physical, the difference can be attributed to electrokinetic processes (not accounted by diffusion coefficients) which are supposed to affect the transport numbers of each ion \cite{Stokes}. In any case, these electrokinetic processes were predicted in the framework of classical, continuum theory of electrolytes and are not clearly observed in our simulations, so its molecular origin remains unclear.

The effect of the damping constant of the Langevin thermostat was investigated by performing molecular dynamics simulations on the same system of size $L = 7.73$nm and same conditions but using a Langevin thermostat with a damping constant of $ \tau_{\text{Lan}} = 0.2$ ps$^{-1}$ (instead of the damping constant of $1$ ps$^{-1}$ employed in all other simulations). The electrolyte exhibits an ohmic behaviour, with conductivity $\kappa = 13.5$ S/m, which slightly differs from the value obtained before and from the experimental value. Nevertheless, the major effect of the damping constant is in the different ionic contributions to the total current, as can be seen in Fig. \ref{Table:IRatio_MgCl2_02}. In comparison to the results of the simulations with $ \tau_{\text{Lan}} = $ ps$^{-1}$, it is evident that there is a spurious dependence of the ratio of the anionic current over the cationic current on the magnitude of the applied electric field.

\begin{table}
\caption{Ionic contributions to the total current, ratio of the anionic current over the cationic current and cationic transport numbers for different values of the applied electric field in the case of a 1M MgCl$_2$ electrolyte, $ L =7.73$nm, and using a damping constant of the Langevin thermostat of $0.2$ ps$^{-1}$. All these quantities are calculated in the frame of reference of the moving fluid.}
\label{Table:IRatio_MgCl2_02}
\centering
%\left
%\begin{tabular}{l|c c c | c c}
\begin{tabular}{|c|c|c|c|c|}
\hline \hline
Electric field (V/nm) & I$_{\text{Cl}}$ (nA) & I$_{\text{Mg}}$ (nA) & I$_{\text{Cl}}$/I$_{\text{Mg}}$ & t$_+$ \\
\hline
%\multicolumn{1}{|c|}{\multirow{4}{*}{L = 7.73 nm}} &
 14.2 & 5.88 & 5.25 & 1.12 &  0.47 \\
 25.98 & 11.49 & 9.56 & 1.20 & 0.45 \\
 43.3 & 19.82 & 15.03 & 1.32 & 0.43\\
 86.6 & 38.59 & 31.15 & 1.24 & 0.45 \\
\hline

\end{tabular} 
\end{table}

%\begin{figure}[ht!]
%   \begin{center}
%      \includegraphics[angle=-90,width=7cm]{Iratiow_MgCl4.eps}
%      \includegraphics[angle=-90,width=7cm]{Iratiow_MgCl8.eps}
%      \caption{(a) Ratio of the anionic current over the cationic current for the different electric fields for the case with $L = 3.82$nm and a Langevin damping constant of 1ps$^{-1}$.(b) Ratio of the anionic current over the cationic current for the different electric fields for the case with $L = 7.73$nm and a Langevin damping constant of 1ps$^{-1}$.}
%      \label{Fig:ratioMgCl}
%   \end{center}
% \end{figure}

%\begin{figure}[ht!]
%   \begin{center}
%      \includegraphics[angle=-90,width=8cm]{TransportNumbers_MgCl4.eps}
%      \includegraphics[angle=-90,width=8cm]{TransportNumbers_MgCl8.eps}
%      \caption{Cationic transport numbers of 1M KCl for different electric fields using two different simulation box sizes $L$. The dots are the results of the calculation from the Molecular Dynamics simulations results for (a)$L = 3.82$, and (b) $L = 7.73$nm. The dashed line gives the experimental value of the cationic transport number for a 1M MgCl$_2$ electrolyte \cite{??} }
%      \label{Fig:TransportNumbersMgCl2}
%   \end{center}
% \end{figure}

\subsubsection{Electoosmotic flow}

\begin{table*}
\caption{Hydration of the different ions}
\label{Table:Hydration}
\centering
\begin{tabular}{ccc|c|c|c|}
\cline{4-6}
& & &\multicolumn{3}{|c|}{Number of attached H$_2$O} \\ \cline{1-6}
\multicolumn{1}{|c|}{Electrolyte} & \multicolumn{1}{|c|}{Ion} & Hydration & $\tau =10$ ps & $\tau =100$ ps & $\tau =1000$ ps \\ \cline{1-6}
\multicolumn{1}{|c|}{\multirow{2}{*}{1M KCl}} &
\multicolumn{1}{|c|}{K$^+$}&6.4  & 1.88 & 0.070 & 0.0012     \\ 
\multicolumn{1}{|c|}{}                        &
\multicolumn{1}{|c|}{Cl$^-$}& 7.3 & 2.41 & 0.11 & 0.0033     \\ \cline{1-6}
\multicolumn{1}{|c|}{\multirow{2}{*}{1M MgCl$_2$}} &
\multicolumn{1}{|c|}{Mg$^{2+}$}& 6.0 & 5.41 & 5.35 & 5.31  \\ 
\multicolumn{1}{|c|}{}                        &
\multicolumn{1}{|c|}{Cl$^-$}& 7.3 & 2.99 & 0.19 & 0.005  \\ \cline{1-6}
\multicolumn{1}{|c|}{\multirow{3}{*}{1M KCl \& 1M MgCl$_2$}} &
\multicolumn{1}{|c|}{Mg$^{2+}$}&6.0  & 5.33& 5.22  & 5.13   \\ 
\multicolumn{1}{|c|}{}                        &
\multicolumn{1}{|c|}{K$^+$}& 6.9 & 1.95&  0.11 &  0.003 \\ 
\multicolumn{1}{|c|}{}                        &
\multicolumn{1}{|c|}{Cl$^-$}&7.0 &3.04 &  0.25 &  0.009 \\ \cline{1-6}
\multicolumn{1}{|c|}{\multirow{2}{*}{1M LaCl$_3$}} &
\multicolumn{1}{|c|}{La$^{3+}$}& 8.4 & 6.98 & 4.65 & 0.27  \\ 
\multicolumn{1}{|c|}{}                        &
\multicolumn{1}{|c|}{Cl$^-$}& 6.9 & 3.50 & 0.35 & 0.009  \\ \cline{1-6}

\end{tabular}
\end{table*}

\begin{figure}[ht!]
   \begin{center}
      \includegraphics[angle=-90,width=8.5cm]{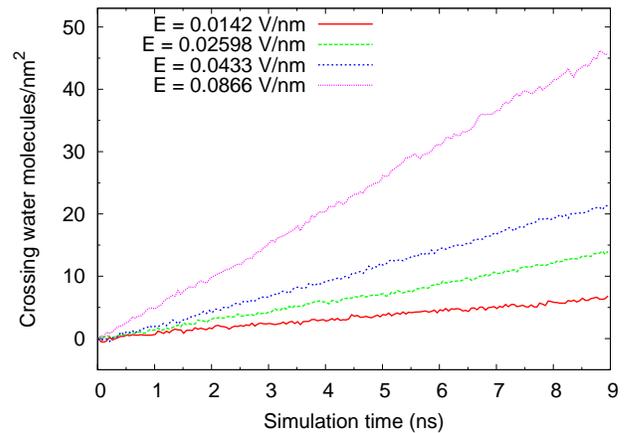}
      \caption{Accumulated number of water molecules per unit area crossing a plane perpendicular to the applied electric field versus simulation time for the 1M MgCl$_2$ electrolyte. Different lines represent the flow of water for different values of the electric field.}
      \label{Fig:water_flowMgCl2}
   \end{center}
 \end{figure}

\begin{figure}[ht!]
   \begin{center}
      \includegraphics[angle=-90,width=9cm]{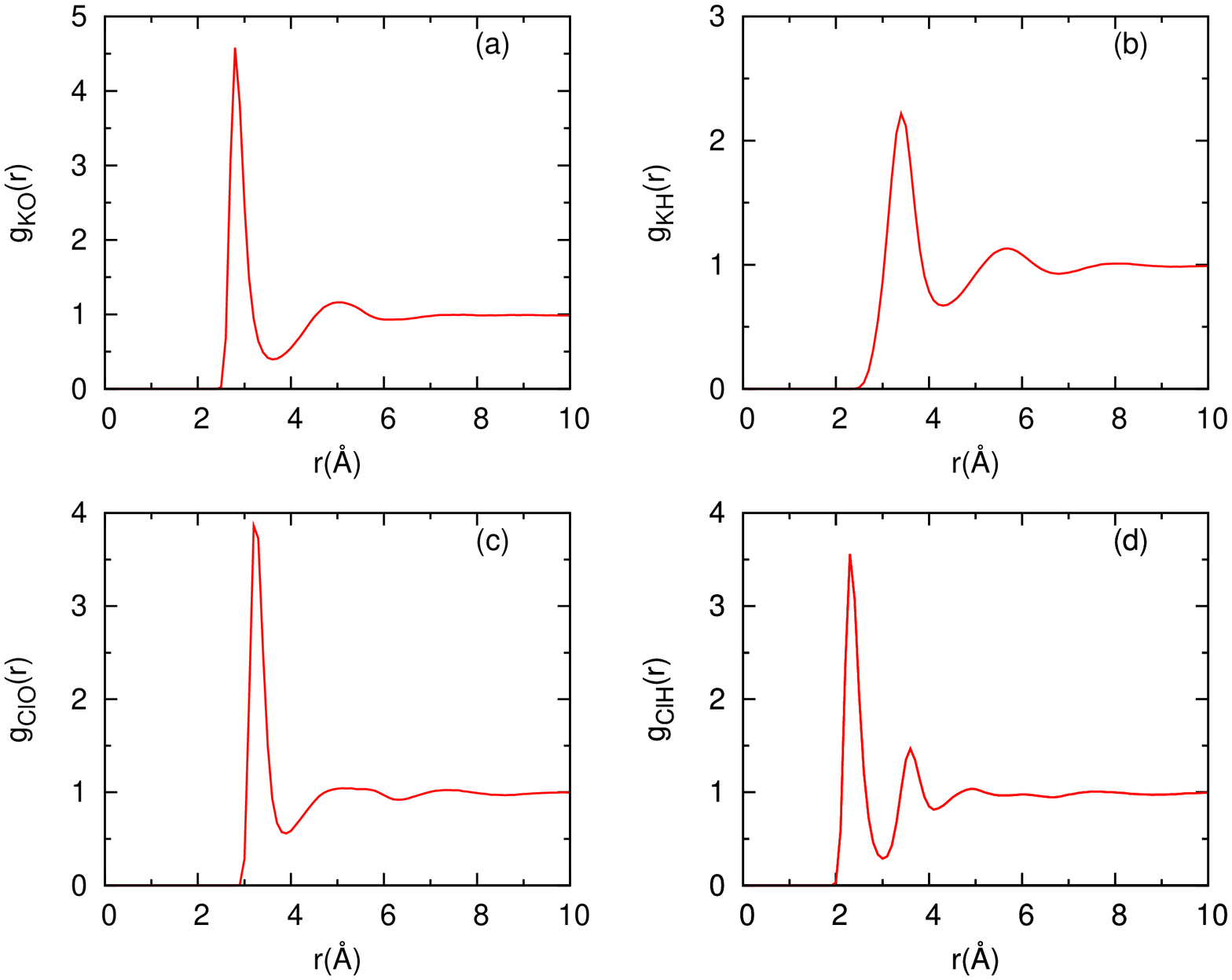}
      \caption{The ion-oxygen and ion-hydrogen radial distribution functions for the 1M KCl electrolyte}
      \label{Fig:multi_RDF_KCl}
   \end{center}
 \end{figure}

\begin{figure}[ht!]
   \begin{center}
      \includegraphics[angle=-90,width=9cm]{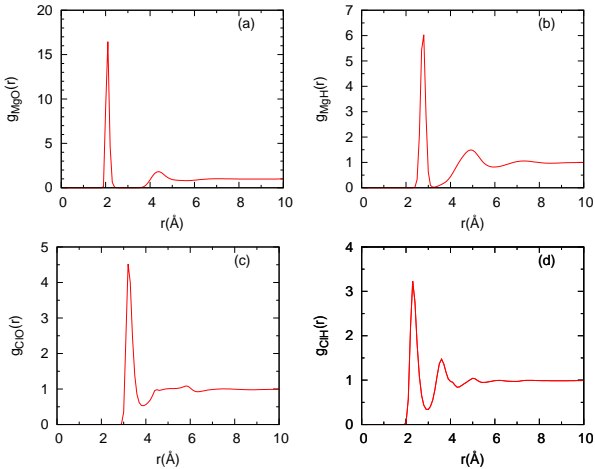}
      \caption{The ion-oxygen and ion-hydrogen radial distribution functions for the 1M MgCl$_2$ electrolyte}
      \label{Fig:multi_RDF_MgCl2}
   \end{center}
 \end{figure}

The electroosmotic flow induced by the ionic current was also computed. In Fig. \ref{Fig:water_flowMgCl2}, the accumulated number of water molecules per unit area crossing a plane perpendicular to the electric field is represented as a function of simulation time. The flux of water molecules for each electric field is given in Table \ref{Table:Water_Flow}. It is interesting to note the different direction and magnitude of the water flow obtained for 1M KCl and for 1M MgCl$_2$ electrolytes. Indeed, while in the presence of 1M KCl the direction of the net flow of water is opposite to the direction of the electric field, in the presence of 1M MgCl$_2$ the net flow of water goes in the direction of the applied field. The magnitude of the net flow of water also differs significantly in both cases, being much larger (almost an order of magnitude) for the 1M MgCl$_2$ electrolyte. These differences can be understood from the hydration properties of the ions involved. For each ion, we have computed its hydration, i.e., the average number of water molecules in its first coordination shell (as defined by the first minimum of the radial distribution function, see Figs. \ref{Fig:multi_RDF_KCl} and \ref{Fig:multi_RDF_MgCl2}). We have also computed the average number of water molecules which have remained a time $\tau$ in the first coordination shell of each ion to test the robustness of the hydration layer, see Table \ref{Table:Hydration}. The values given in Table \ref{Table:Hydration} do not show any dependence on the value of the electric field applied in the simulations. The reported results for the hydration values of the different ions agree with results from previous molecular dynamics studies  \cite{Chowdhuri, Callahan, Jiao} and Ab initio calculations\cite{Lightstone}. The results for the average number of water molecules that spend a time $\tau$ in the first coordination shell of the different ions are also in agreement with previous computational studies, in which residence times of water molecules in the first coordination shell of the order of $\sim 10$ps were obtained for K$^+$ and Cl$^-$ and of the order of $\sim 10$ns for Mg$^{2+}$. Such difference in the robustness of the hydration layer of Mg$^{2+}$ and the other ions is a broadly accepted fact, established by ab initio and DFT calculations \cite{Lightstone,Waizumi} as well as by NMR, Raman spectroscopy, and X-ray adsorption spectroscopy experiments \cite{Struis1989,Pye,Cappa}.

Hence, from the analysis of the hydration properties of the ions it is possible to understand the difference in the electroosmotic flow between the 1M KCl and 1M MgCl$_2$ electrolytes. For the 1M KCl electrolyte, the anionic current is slightly greater than the cationic current and the hydration of Cl$^-$ is higher than that of K$^+$ so it exhibits a net electroosmotic flow in the direction of the flow of Cl$^-$. For the 1M MgCl$_2$ electrolyte, although the current of Cl$^-$ is greater than the current of Mg$^{2+}$, the latter ion is much more effective dragging water molecules (its hydration layer is much more robust over time) so the net electroosmotic flow is in the direction of the flow of Mg$^{2+}$ along the applied electric field.

\newpage

\subsubsection{Concentration dependence of transport properties of MgCl$_2$}

\begin{figure}[ht!]
   \begin{center}
      \includegraphics[angle=-90,width=8.5cm]{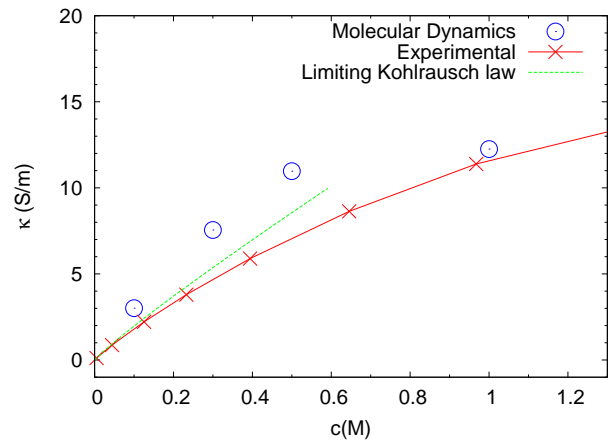}
      \caption{Electric conductivity of a MgCl$_2$ electrolyte versus its salt concentration for an electric field $E = 43.3$mV/nm. Blue dots represent results obtained from MD simulations, red crosses experimental data \cite{Phang1980}, and the green dashed line is Kohlrausch's limiting law \cite{Stokes}.}
      \label{Fig:cond_vs_conc}
   \end{center}
 \end{figure}

\begin{figure}[ht!]
   \begin{center}
      \includegraphics[angle=-90,width=8.5cm]{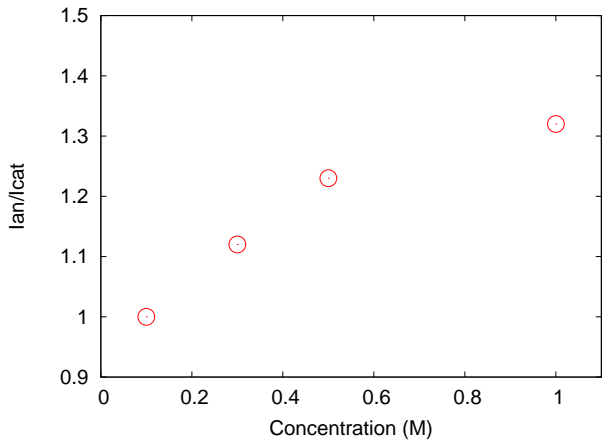}
      \caption{Ratio of the contribution to the electric current of a MgCl$_2$ electrolyte by anions and cations versus its bulk concentration when an electric field $E = 43.3$mV/nm is applied.}
      \label{Fig:ratio_vs_conc}
   \end{center}
 \end{figure}

For the MgCl$_2$ electrolyte we have also tested the dependency of the conduction properties of the electrolyte on its concentration, $c$. We have performed simulations at different concentrations of MgCl$_2$ for a fixed value of the applied electric field $E = 0.0433$V/nm in a cubic simulation box of side $L \simeq 8$nm (see Table \ref{Table:Table1} for precise values). The summary of the results are given in Table \ref{Table:cond_vs_conc} and Figures \ref{Fig:cond_vs_conc} and \ref{Fig:ratio_vs_conc}. Here, for each value of the electrolyte concentration the conductivity has been obtained from a single point $(I, E)$, under the assumption that the electrolyte exhibits an ohmic behavior (suggested by the studies at 1M concentration, see Fig. \ref{Fig:MgCl}). As expected, the electrical conductivity, $\kappa$, of the MgCl$_2$ solution increases with increasing concentration, as can be seen in Table \ref{Table:cond_vs_conc} and Fig. \ref{Fig:cond_vs_conc}. In Fig. \ref{Fig:cond_vs_conc}, the conductivity of the electrolyte obtained from molecular dynamics simulations for each concentration is compared with experimental results \cite{Phang1980} and the Kohlrausch's limiting law. The results obtained from MD simulations reproduce well the dependency of the electrolyte conductivity on the concentration, as compared to the experimental tendency. Besides, the absolute values of the experimental and simulated conductivities agree well with each other, especially for high and low salt conditions.
In Fig. \ref{Fig:cond_vs_conc} we represent the ratio of the anionic current over the cationic current as a function of the MgCl$_2$ concentration. The results from MD simulations indicate that the contribution of the anions becomes more prominent with increasing concentration of electrolyte. This tendency agrees with the experimental behaviour for $2:1$ electrolytes \cite{Stokes}.

\begin{table}
\caption{Total current, electrical conductivity, ionic contributions to the total current, ratio of the anionic current over the cationic current and cationic transport numbers for different values of the concentration of the MgCl$_2$ electrolyte.}
\label{Table:cond_vs_conc}
\centering
%\left
%\begin{tabular}{l|c c c | c c}
    \begin{tabular}{|c|cccccc|}
\hline
 c(M)& Itot (nA)& $\kappa$(S/m) & I$_{\text{Mg}}$ (nA) & I$_{\text{Cl}}$ (nA)& Ian/Icat & t$_+$  \\
\hline
 0.1 & 8.0  & 3.01 & 4.0 & 4.0 & 1.00 & 0.5 \\
 0.3 & 19.97 & 7.55 & 9.42 & 10.54 & 1.12 & 0.47\\
 0.5 & 28.79 & 10.97 & 12.90 & 15.88 & 1.23 & 0.45\\
 1.0 & 31.70 & 12.25 & 13.66 & 18.05 & 1.32 & 0.43\\

\hline
    
    \end{tabular} 
\end{table}

\subsection{Transport properties of mixture composed of 1M KCl and 1M MgCl$_2$}
We have also studied the transport properties of an electrolyte composed of 1M KCl and 1M MgCl$_2$. The addition of monovalent ions to a system immersed in a multivalent electrolyte is sometimes used experimentally to screen the electric charge of the multivalent ions \cite{Heyden2006}. This effect is also used in experiment and simulations to determine whether electrostatic correlations are responsible for a certain macroscopic effect \cite{Alberto, MarcelOmpF, AlbertoPRL}.

\begin{figure}[ht!]
   \begin{center}
      \includegraphics[angle=-90,width=8.5cm]{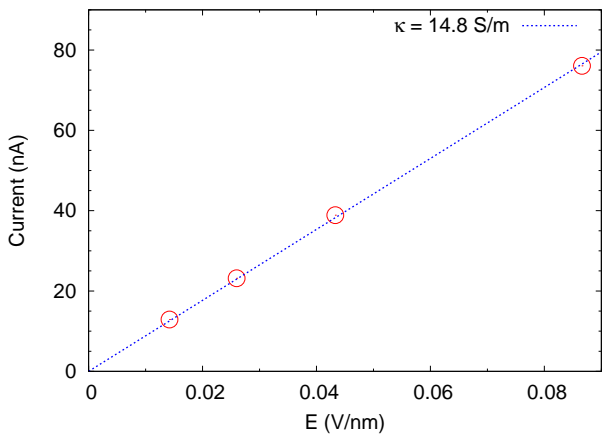}
      \caption{MD results for the electric current versus applied electric field for an electrolyte composed of 1M KCl and 1M MgCl$_2$ in a cubic simulation box of $L = 7.72$nm. The points correspond to simulation data computed from Eq.(\ref{Ieq}) and lines are fits of the data to Eq.(\ref{conductivity}) (fit gives $\kappa = 14.8$S/m)}
      \label{Fig:KClMgCl8}
   \end{center}
 \end{figure}

The analysis is similar to the analysis done for the previous electrolytes. The electric currents caused by the drift of ions have been obtained for four different values of the applied electric field. In this case, a single cubic box of size $L = 7.72$ nm was employed. The results for the conductivity are summarized in Fig. \ref{Fig:KClMgCl8}, which results in a value for the conductivity $\kappa = 14.8$ S/m. Our own experimental measurements give $\kappa = 16.83 \pm 0.01$ S/m. The different contribution to the total current of the different ions, as well as the cationic transport number are given in Table \ref{Table:IRatio_KClMgCl2} for every value of the applied electric field. 

%\begin{figure}[ht!]
%   \begin{center}
%      \includegraphics[angle=-90,width=8.5cm]{Iratiow_KClMgCl8.eps}
%      \caption{Ratio of the anionic current over the cationic current for the different electric fields for the case with electrolyte composed of 1M KCl and 1M MgCl$_2$, with $L = 7.72$nm and a Langevin damping constant of 1ps$^{-1}$.}
%      \label{Fig:ratiow_KClMgCl8}
%   \end{center}
% \end{figure}

%\begin{figure}[ht!]
%   \begin{center}
%      \includegraphics[angle=-90,width=8.5cm]{Iratio2w_KClMgCl8.eps}
%      \caption{Ratio of the different contributions to the current over the total current for the case with electrolyte composed of 1M KCl and 1M MgCl$_2$, with $L = 7.72$nm and a Langevin damping constant of 1ps$^{-1}$.}
%      \label{Fig:ratio2w_KClMgCl8}
%   \end{center}
% \end{figure}

\begin{table}
\caption{Ionic contributions to the total current, ratio of the anionic current over the cationic current and cationic transport numbers for different values of the applied electric field in the case of an electrolyte composed by 1M MgCl$_2$ and 1M KCl. All these quantities are calculated in the frame of reference of the moving fluid.}
\label{Table:IRatio_KClMgCl2}
\centering
%\left
%\begin{tabular}{l|c c c | c c}
\begin{tabular}{|c|c|c|c|c|c|}
\hline \hline
 Electric field (V/nm) & I$_{\text{Cl}}$(nA) & I$_{\text{K}}$(nA) & I$_{\text{Mg}}$(nA) & I$_{\text{an}}$/I$_{\text{cat}}$ & t$_+$ \\
\hline
14.2 & 7.6 & 3.27 & 2.04 & 1.43 & 0.41 \\
25.98 & 18.14 & 5.12 & 4.98 & 1.30 & 0.43 \\
43.3 & 21.69 & 9.29 & 7.89 & 1.26 & 0.44\\
86.6 & 43.36 & 17.11 & 15.59 & 1.32 & 0.43 \\
\hline

\end{tabular} 
\end{table}

\subsubsection{Electroosmotic flow}
 
\begin{figure}[ht!]
   \begin{center}
      \includegraphics[angle=-90,width=8.5cm]{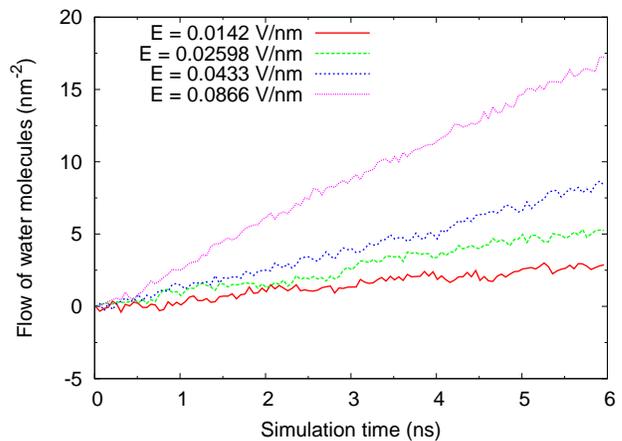}
      \caption{Accumulated number of water molecules per unit area crossing a plane perpendicular to the applied electric field versus simulation time for an electrolyte composed of 1M MgCl$_2$ and 1M KCl. Different lines represent the flow of water for different values of the electric field.}
      \label{Fig:water_flowKClMgCl2}
   \end{center}
 \end{figure}

The results for the net flow of water induced by the ionic current are given in Fig. \ref{Fig:water_flowKClMgCl2}, in which the accumulated number of water molecules per unit area crossing a plane perpendicular to the electric field is represented as a function of simulation time. The flux of water molecules for each electric field is given in Table \ref{Table:Water_Flow}. Similarly to the case of the 1M MgCl$_2$ electrolyte, the net electroosmotic flow is in the direction of the flow of cations, in spite of being smaller than the net flow of anions (see Table \ref{Table:IRatio_KClMgCl2}). We interpret this result along the same lines as with the case of 1M MgCl$_2$ electrolyte. As it is shown in Table \ref{Table:Hydration}, the hydration layer of Mg$^{2+}$ is much more robust than the hydration of Cl$^-$ and K$^+$, so the electroosmotic flow induced by the flow of Mg$^{2+}$ dominates. 

\subsection{Transport properties of 1M LaCl$_3$}
The lanthanum cation La$^{3+}$ is a highly charged and polarizable ion which, for most problems, requires the use of Ab initio calculations or polararizable force fields to properly describe its interaction with water and ions \cite{Clavaguera}. However, there are systems for which only its electric charge and size are relevant to describe certain electrokinetic phenomena \cite{AlbertoPRL}. For such cases, La$^{3+}$ can be modeled as a charged Lennard-Jones particle and one can obtain the transport properties of a LaCl$_3$ ionic solution by using classical molecular dynamics calculations. The Lennard-Jones parameters used here to describe the lanthanum cation La$^{3+}$ were taken from Ref.\cite{Veggel}

\begin{figure}[ht!]
   \begin{center}
      \includegraphics[angle=-90,width=8.5cm]{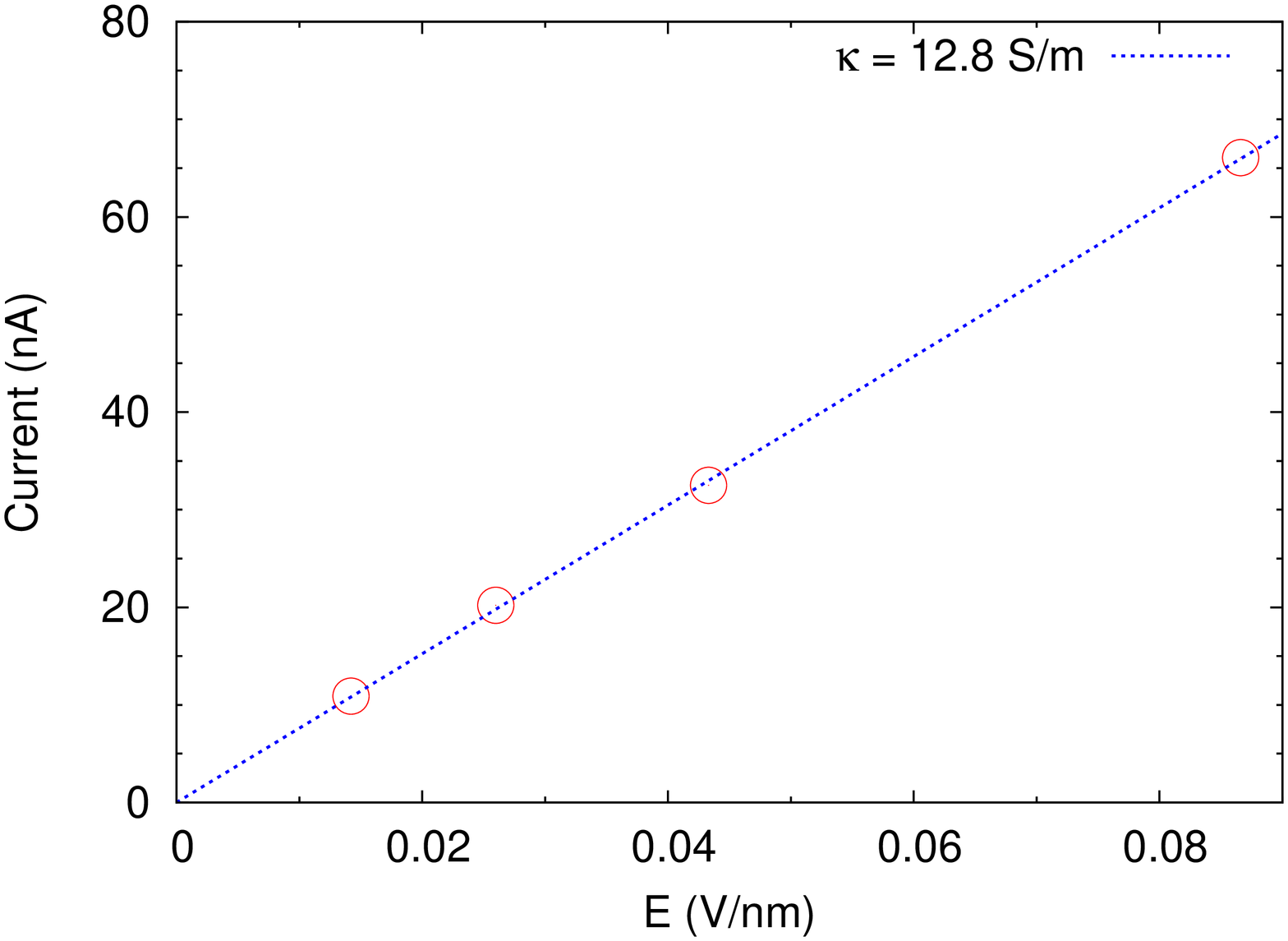}
      \caption{MD results for the electric current versus applied electric field for a 1M LaCl$_3$ electrolyte in a cubic simulation box of $L = 7.71$nm. The points correspond to simulation data computed from Eq.(\ref{Ieq}) and lines are fits of the data to Eq.(\ref{conductivity}) (fit gives $\kappa = 12.8$S/m)}
      \label{Fig:LaCl8}
   \end{center}
 \end{figure}

The electrical transport properties of the 1M LaCl$_3$ electrolyte in the presence of an external electric field are obtained using molecular dynamics simulations which employ a non-polarizable force field (described above). The summary of the results is given in Fig.\ref{Fig:LaCl8} and Table \ref{Table:IRatio_LaCl3}. The electrolyte shows an ohmic behaviour (see Fig.\ref{Fig:LaCl8}), with an electrical conductivity of $\kappa = 12.8$ S/m. Considering the limitations of the description, this value is quite satisfactory compared to the experimental value of $\kappa_\text{{exp}} = 15.3$ S/m \cite{Grinnell}. The ionic contribution, the ratio of currents and the cationic transport number are given in Table \ref{Table:IRatio_LaCl3} for each value of the external electric field.

\begin{table}
\caption{Ionic contributions to the total current, ratio of the anionic current over the cationic current and cationic transport numbers for different values of the applied electric field in the case of the 1M LaCl$_3$ electrolyte, with $L = 7.71$nm. All these quantities are calculated in the frame of reference of the moving fluid.}
\label{Table:IRatio_LaCl3}
\centering
%\left
%\begin{tabular}{l|c c c | c c}
\begin{tabular}{|c|c|c|c|c|}
\hline \hline
Electric field (V/nm) & I$_{\text{Cl}}$ (nA) & I$_{\text{La}}$ (nA) & I$_{\text{Cl}}$/I$_{\text{La}}$ & t$_+$ \\
\hline
 14.2 & 5.51 & 5.00 & 1.10 &  0.48 \\
 25.98 & 10.72 & 9.89 & 1.13 & 0.47 \\
 43.3 & 17.55 & 14.94 & 1.17 & 0.46\\
 86.6 & 35.00 & 31.06 & 1.12 & 0.47 \\
\hline
\end{tabular} 
\end{table}

%\begin{figure}[ht!]
%   \begin{center}
%      \includegraphics[angle=-90,width=8.5cm]{Iratiow_LaCl8.eps}
%      \caption{Ratio of the anionic current over the cationic current for the different electric fields for a 1M LaCl$_3$ electrolyte, with $L = 7.71$nm and a Langevin damping constant of 1ps$^{-1}$.}
%      \label{Fig:RatioLaCl8}
%   \end{center}
% \end{figure}

\subsubsection{Electroosmotic flow}
 
\begin{figure}[ht!]
   \begin{center}
      \includegraphics[angle=-90,width=8.5cm]{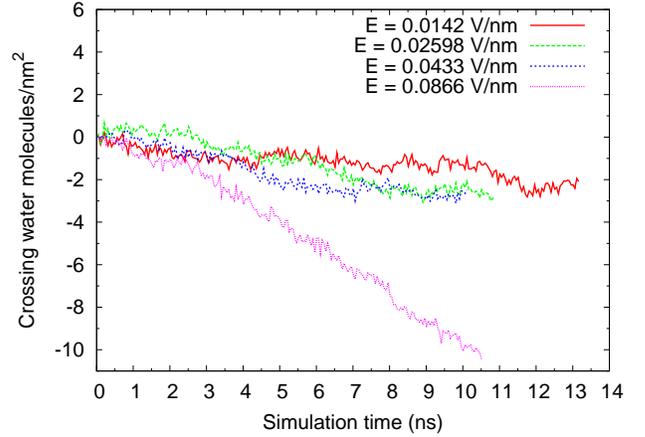}
      \caption{Accumulated number of water molecules per unit area crossing a plane perpendicular to the applied electric field versus simulation time for an electrolyte composed of 1M LaCl$_3$. Different lines represent the flow of water for different values of the electric field.}
      \label{Fig:water_flowLaCl3}
   \end{center}
 \end{figure}

The results for the net flow of water induced by the ionic current are given in Fig. \ref{Fig:water_flowLaCl3}, in which the accumulated number of water molecules per unit area crossing a plane perpendicular to the electric field is represented as a function of simulation time. The flux of water molecules for each electric field is given in Table \ref{Table:Water_Flow}. 

The hydration properties of the 1M LaCl$_3$ electrolyte are summarized in Table \ref{Table:Hydration}. In spite of the limitations of the non-polarizable model used to describe La$^{3+}$, the values obtained for the hydration of La$^{3+}$ are close to the values given in the literature (hydration $\sim 8-9$ and a residence time of hydrated water in the first coordination shell of $\sim 1$ns \cite{Clavaguera}). As showed in Fig. \ref{Fig:water_flowLaCl3}, the electroosmotic flow obtained for the 1M LaCl$_3$ electrolyte is in the direction of the flow of chloride ions, opposite to the direction of the applied electric field. In this case, the higher hydration of La$^{3+}$ is not enough to compensate the higher flow of chloride ions versus the flow of lanthanum cation La$^{3+}$. 

\begin{figure}[ht!]
   \begin{center}
      \includegraphics[angle=-90,width=9cm]{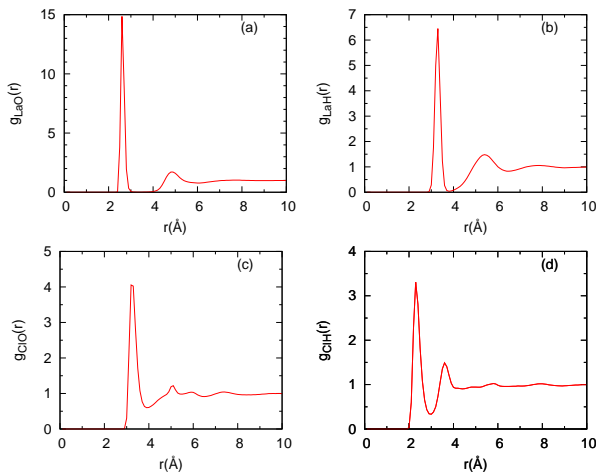}
      \caption{The ion-oxygen and ion-hydrogen radial distribution functions for the 1M LaCl$_3$ electrolyte}
      \label{Fig:multi_RDF_LaCl3}
   \end{center}
 \end{figure}

\section{Conclusions}

Classical molecular dynamics simulations of various electrolytes (KCl, MgCl$_2$, KCl + MgCl$_2$, LaCl$_3$) in external electric fields were performed to study their transport properties. We have employed algorithms and parameters typically used in simulations of complex electrokinetic phenomena to test against experimental data their validity in the description of transport properties of electrolytes. The electrical conductivity and transport numbers (ionic contributions to the total current) of electrolytes containing monovalent (1M KCl), divalent (1M MgCl$_2$, 1M MgCl$_2$ + 1M KCl ), and  trivalent (1M LaCl$_3$) cations were computed from the simulated ion trajectories. It is shown that in all cases the electrical conductivities obtained from the simulations are in good agreement with our own experimental measurements and values from the literature. Also, for the MgCl$_2$ electrolyte, the dependence of the electrical conductivity on the electrolyte concentration was investigated and found that the MD results follow the experimental trend and give accurate results for the lowest (0.1M) and the highest concentrations (1M) studied. Transport numbers obtained for the 1M KCl electrolyte agree well with the experimental values found in the literature. For the divalent electrolyte 1M MgCl$_2$, there is a significant discrepancy between the transport numbers obtained from MD simulations and the experimental values, probably due to the poor result for the self difusion coefficient for anions provided by the Lennard-Jones parameters used in the simulations. The effect of the simulation box size on the electrical transport properties was also explored. It is found that a big enough simulation box is needed to avoid spurious effects of the simulation.

The electroosmotic flow of water induced by the ionic flow has also been computed from the MD trajectories. It is shown that the direction and magnitude of the water flux depends on the electrolyte: while the flux is in the direction of the cation flow for electrolytes containing Mg$^{2+}$, it is in the opposite direction and weaker for all the other cases. These results are interpreted with the help of the hydration properties of the ions, which are calculated and successfully compared with previous studies on the subject. 

\section{Acknowledgments}
This work is supported by the Spanish Government (grants FIS2009-13370-C02-02, FIS2007-60205 and CONSOLIDER-NANOSELECT-CSD2007-00041), Generalitat de Catalunya (2009SGR164) and Fundaci\'o Caixa Castell\'o-Bancaixa (P1-1A2009-13). C.C. is supported by the JAE doc program of the Spanish National Research Council (CSIC). The Supercomputing resources employed in this work were provided by the CESGA Supercomputing Center, Spain.

\end{document}